\documentclass[aps,prl,twocolumn,superscriptaddress]{revtex4-1}

\usepackage{graphicx}
\usepackage{dcolumn}
\usepackage{bm}
\usepackage{color}
\usepackage{amsmath}

\begin{document}

\title{
Quantum dynamics of hydrogen in iron-based superconductor LaFeAsO$_{0.9}$D$_{0.1}$ measured with inelastic neutron spectroscopy
}

\author{Jun-ichi~Yamaura}
\email{jyamaura@mces.titech.ac.jp}
\affiliation{Materials Research Center for Element Strategy, Tokyo Institute of Technology, Yokohama, Kanagawa 226-8503, Japan}

\author{Haruhiro~Hiraka}
\email{hiraka@kaeri.re.kr}
\altaffiliation{Korea Atomic Energy Research Institute at present.}
\affiliation{Condensed Matter Research Center, Institute of Materials Structure Science, KEK, Tsukuba, Ibaraki~305-0801, Japan}

\author{Soshi~Iimura}
\affiliation{Laboratory for Materials and Structures, Tokyo Institute of Technology, Yokohama, Kanagawa 226-8503, Japan}

\author{Yoshinori~Muraba}
\affiliation{Materials Research Center for Element Strategy, Tokyo Institute of Technology, Yokohama, Kanagawa 226-8503, Japan}

\author{Joonho~Bang}
\affiliation{Materials Research Center for Element Strategy, Tokyo Institute of Technology, Yokohama, Kanagawa 226-8503, Japan}

\author{Kazuhiko~Ikeuchi}
\affiliation{Comprehensive Research Organization for Science and Society, Tokai, Ibaraki~319-1106, Japan}

\author{Mitsutaka~Nakamura}
\affiliation{J-PARC Center, Japan Atomic Energy Agency, Tokai, Ibaraki~319-1195, Japan}

\author{Yasuhiro~Inamura}
\affiliation{J-PARC Center, Japan Atomic Energy Agency, Tokai, Ibaraki~319-1195, Japan}

\author{Takashi~Honda}
\affiliation{Condensed Matter Research Center, Institute of Materials Structure Science, KEK, Tsukuba, Ibaraki~305-0801, Japan}

\author{Masatoshi~Hiraishi}
\affiliation{Condensed Matter Research Center, Institute of Materials Structure Science, KEK, Tsukuba, Ibaraki~305-0801, Japan}

\author{Kenji~M.~Kojima}
\affiliation{Condensed Matter Research Center, Institute of Materials Structure Science, KEK, Tsukuba, Ibaraki~305-0801, Japan}

\author{Ryosuke~Kadono}
\affiliation{Condensed Matter Research Center, Institute of Materials Structure Science, KEK, Tsukuba, Ibaraki~305-0801, Japan}

\author{Yoshio~Kuramoto}
\affiliation{Condensed Matter Research Center, Institute of Materials Structure Science, KEK, Tsukuba, Ibaraki~305-0801, Japan}
\affiliation{Department of Physics, Kobe University, Kobe, Hyogo 657-8501, Japan}

\author{Youichi~Murakami}
\affiliation{Materials Research Center for Element Strategy, Tokyo Institute of Technology, Yokohama, Kanagawa 226-8503, Japan}
\affiliation{Condensed Matter Research Center, Institute of Materials Structure Science, KEK, Tsukuba, Ibaraki~305-0801, Japan}

\author{Satoru~Matsuishi}
\affiliation{Materials Research Center for Element Strategy, Tokyo Institute of Technology, Yokohama, Kanagawa 226-8503, Japan}

\author{Hideo~Hosono}
\affiliation{Materials Research Center for Element Strategy, Tokyo Institute of Technology, Yokohama, Kanagawa 226-8503, Japan}
\affiliation{Laboratory for Materials and Structures, Tokyo Institute of Technology, Yokohama, Kanagawa 226-8503, Japan}

\date{\today}
\begin{abstract}
Inelastic neutron scattering was performed for an iron-based superconductor LaFeAsO$_{0.9}$D$_{0.1}$, where most of D (deuterium) replaces oxygen, while a tiny amount goes into interstitial sites. By first-principle calculation, we characterize the interstitial sites for D (and for H slightly mixed) with four equivalent potential minima. Below the superconducting transition temperature $T_{\rm c} = 26$~K, new excitations emerge in the range 5--15 meV, while they are absent in the reference system LaFeAsO$_{0.9}$F$_{0.1}$. The strong excitations at 14.5~meV and 11.1~meV broaden rapidly around 15~K and 20~K, respectively, where each energy becomes comparable to twice of the superconducting gap. The strong excitations are ascribed to a quantum rattling, or a band motion of hydrogen, which arises only if the number of potential minima is larger than two.
\end{abstract}

\maketitle

The discovery of iron-based superconductors has stimulated a growing body of research~\cite{kamihara,paglione,stewart,hosono}. For identifying the mechanism of superconductivity, one needs detailed information of the superconducting gap, together with its temperature dependence. In addition to thermodynamic and transport studies, direct measurements of the energy gaps have been tried by various methods such as photoemission~\cite{Ding2008,Kondo2008}, tunneling~\cite{Hanaguri2010}, point-contact~\cite{Chen2008,gonnelli09}, and infrared~\cite{Li2008,Gorshunov2010} spectroscopies. These methods require high quality of the sample, which is not feasible in some iron-based superconductors made with intricate procedure of synthesis. Therefore, a new direct method is awaited that is also applicable to polycrystalline samples with rough surfaces, magnetic impurities, and other imperfections.

Here, we propose a new method for direct detection of the superconducting gap in the iron-based superconductor via coupling to hydrogen motion at interstitial sites. When two or more interstitial sites of hydrogen are close, the overlap of wave functions causes
the splitting of hydrogen energy levels~\cite{fukai,kehr78}. This quantum phenomenon is known as the tunneling splitting, and has long been studied in conventional superconductors by inelastic neutron scattering (INS) ~\cite{fukai,wipf97,kondo86,kagan92}. The magnitudes of the tunneling splitting in known cases are much smaller than the characteristic value of the superconducting gap. As we shall demonstrate, the hydrogen vibration in the iron-based superconductors is highly anharmonic at the interstitial sites. Accordingly, the generalized tunneling splitting has exceptionally large energy of the order of 10 meV, which is comparable to the superconducting gap. Hence, the hydrogen motion serves as a good probe for superconductivity. Since the local dynamics is not much influenced by the sample quality, the hydrogen probe has a unique advantage over the conventional spectroscopies.

We performed INS measurements for a deuterated LaFeAsO$_{0.9}$D$_{0.1}$ with $T_{\rm c} = 26$~K~\cite{iimura12}. We found strong intensities in the excitation range of 4--15~meV below $T_{\rm c}$. The source of these excitations is ascribed to hydrogen since the reference material LaFeAsO$_{0.9}$F$_{0.1}$, with the same electronic/magnetic phase diagram~\cite{kamihara,iimura12}, does not show the corresponding intensities. Throughout this paper, we use the term ``hydrogen" for both $^1$H and $^2$H, and protium (H) or deuterium (D) to distinguish the isotopes explicitly.

We prepared a powder sample of LaFeAsO$_{0.9}$D$_{0.1}$ (D-sample), weighing $\sim$30~g, under high pressure~\cite{iimura12}. Hydrogen anions (D$^-$) were substituted for O$^{2-}$ anions~\cite{matsuishi,hiraishi14}. In addition, owing to a tiny amount of H in deuterated reagents~\cite{iimura13}, some mixture of H at the hydrogen sites is inevitable. Their cross sections $\sigma_{\rm inc}$ of incoherent scattering are $\sigma_{\rm inc}$(H)$ = 80.3\times10^{-24}$ cm$^2$ and $\sigma_{\rm inc}$(D)$ = 2.1\times10^{-24}$ cm$^2$. The H/D concentration ratio was estimated to be $\sim$0.02 based on the analysis of excitations at 70--130~meV of H and D confined in the O sites. We also prepared LaFeAsO$_{0.9}$F$_{0.1}$ (F-sample) with $\sim$20~g under ambient pressure as the reference for the D-sample.

The INS intensities for D- and F-samples were measured on a chopper spectrometer (4SEASONS)~\cite{4seasons} at BL01 in the pulsed-neutron source at the Japan Proton Accelerator Research Complex (J-PARC) in the Materials and Life Science Experimental Facility (MLF). Data were collected at incident neutron energies of $E_{\rm i}$ = 9.3, 17.5, and 44.5~meV~\cite{nakamura09}. The Fermi-chopper frequency was set to 150~Hz, with a resolution $\hbar\Delta \omega /E_{\rm i}$ of $\sim$2\% at the energy transfer $\hbar\omega \approx 14$~meV. The time-of-flight data from position-sensitive detectors were converted into the $\bm Q$-angle average $\langle S(\bm {Q},\hbar\omega)\rangle _{\rm angle}$ of the dynamical structure factor by the Utsusemi program~\cite{inamura}. We use the notation $S(Q, \hbar\omega)$ for the angle average.

\begin{figure}
\includegraphics[width=7.5cm]{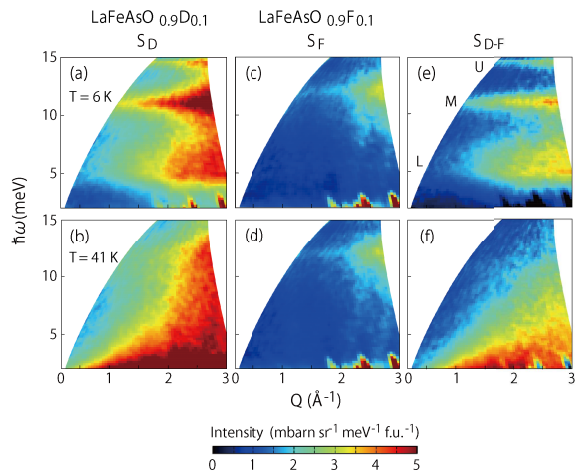}
\caption{\label{fig1}
Color plots of $S_{\rm A}(Q, \hbar\omega)$ with A = D, F, and D--F at 6~K (a,c,e), and 41~K (b,d,f). The incident neutron energy is $E_{\rm i}$ = 17.5~meV. (a, b) $S_{\rm D}$ from LaFeAsO$_{0.9}$D$_{0.1}$; (c, d)  $S_{\rm F}$ from LaFeAsO$_{0.9}$F$_{0.1}$; (e, f) the difference map $S_{\rm D-F}$. U, M, and L in (e) signify the upper, middle, and lower modes, respectively.
}
\end{figure}

Figure \ref{fig1} shows $S(Q, \hbar\omega)$ for D- and F-samples, written as $S_{\rm D}$ and $S_{\rm F}$ respectively, together with their difference $S_{\rm D-F}$. Comparing $S_{\rm D}$ at 6~K below $T_{\rm c}$ (Fig.~\ref{fig1}(a)) and at 41~K above $T_{\rm c}$ (Fig.~\ref{fig1}(b)), we find three groups of excitations at 6~K, all of which are insensitive to $Q$ but merge into a broad feature at 41~K with a strong diffusive character over the entire energy range. In sharp contrast to $S_{\rm D}$, the results $S_{\rm F}$ for the F-sample exhibit neither the flat excitations at 6~K (Fig.~\ref{fig1}(c)) nor the diffuse scattering at 41~K (Fig.~\ref{fig1}(d)). Instead, a small temperature-insensitive feature is visible at $\hbar \omega$ $\approx$ 12 meV in Figs.~\ref{fig1}(c) and \ref{fig1}(d). This is ascribed to phonon scattering as previously reported for the F-sample~\cite{christianson08}. On the other hand, the magnetic scattering reported at $Q \approx$ 1.1~\AA~\cite{shamoto10, wakimoto10} is not visible at the present scale because of their weak intensities.

Figures~\ref{fig1}(e) and \ref{fig1}(f) display the difference map $S_{\rm D-F}\equiv S_{\rm D}-S_{\rm F}$ at 6~K and 41~K, respectively. At 6~K, the three distinct features are assigned as U, M, and L, as indicated in Fig.~\ref{fig1}(e). The D- and F-samples have almost the same electric and superconducting properties~\cite{iimura12}. Consequently, the strong excitations in $S_{\rm D-F}$ must originate from hydrogen.

\begin{figure}
\includegraphics[width=6.3cm]{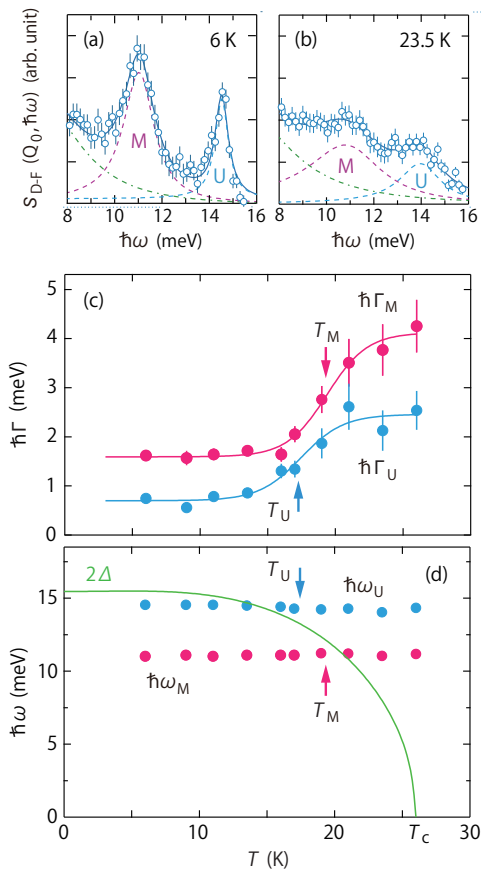}%
\caption{\label{fig2}
Scattering intensities with incident neutron energy $E_{\rm i} = 17.5$~meV at (a) 6~K and (b) 23.5~K for $Q_0 = 2.15\pm 0.15$~\AA$^{-1}$. The indigo solid line represents the sum of all components, the blue dashed line the U component, the red dashed line the M component, and the green dash-dotted line the tail of the L component which works as a base line for M and U components. Also shown are the temperature dependence of (c) the linewidth $\hbar\Gamma_\alpha$ with $\alpha =$ U, M and (d) the peak position $\hbar\omega_\alpha$ for the U and M modes. The blue and magenta solid lines are the results on fitting the U and M peaks, respectively. The green solid line shows the temperature dependence of the BCS gap function 2$\Delta(T)$. The arrows indicate the temperature $T_\alpha$ corresponding to the midpoint of $\hbar\Gamma_\alpha(T)$ as fitted by the sigmoid function.
 }
\end{figure}

Figures~\ref{fig2}(a) and \ref{fig2}(b) display the energy spectra of $S_{\rm D-F}$ at 6~K and 23.5~K, respectively, with $Q$ = 2.15 $\pm$ 0.15~\AA$^{-1}$. For each temperature, the scattering peaks are fitted by a Lorentzian spectrum with the linewidth $\hbar\Gamma_\alpha(T)$ where $\alpha$ represents either U or M mode. Figure~\ref{fig2}(c) plots the temperature dependence of $\hbar\Gamma_\alpha(T)$, which shows that both $\hbar\Gamma_{\rm U}$ and $\hbar\Gamma_{\rm M}$ barely depend on temperature for $T<15$ K. A rapid increase is seen however around 20 K, and then both linewidths are weakly temperature-dependent as the temperature approaches $T_{\rm c} \sim 26$~K. In contrast, each peak position $\hbar\omega_\alpha$ is nearly constant against temperature, as shown in Fig.~\ref{fig2}(d).

It is likely that the rapid increase of $\hbar\Gamma_\alpha$ at $T\lesssim T_{\rm c}$ is related to the coupling of the hydrogen motion to electronic excitations. Namely, with increasing temperature toward $T_{\rm c}$, the superconducting gap $\Delta (T)$ decreases so as to satisfy the condition $\hbar\omega_\alpha > 2\Delta(T)$. This means that $S_{\rm D-F}$ serves as a probe of the superconducting gap. For comparison, we refer to the INS spectra in Nb(OH)$_x$ ($x \approx$ 0.002), where an inelastic peak at $\sim$0.2~meV below $T_{\rm c}$ = 9.2~K merge into the broad quasi-elastic feature above $T_{\rm c}$~\cite{magerl86}. The origin of the peak has been assigned to the tunneling splitting of the interstitial hydrogen in the double well potential. In contrast to the present case, the inelastic peak in Nb(OH)$_x$ is much smaller than $2\Delta_0\sim 3$~meV in the system. The peak width below $T_{\rm c}$ decreases toward the resolution limit upon cooling~\cite{Black79,magerl86}. The change in width is interpreted in terms of a Korringa-type relaxation due to the conduction electrons, which becomes active as the superconducting gap decreases~\cite{wipf97,kondo86,magerl83,magerl86}.

It appears from Fig.~\ref{fig2}(c) that each peak width saturates at both low and high temperatures. We then use the sigmoid function to fit $\hbar\Gamma_\alpha(T)$, as presented in Fig.~\ref{fig2}(c). Next, with a trial value for the superconducting gap $\Delta_0$ at zero temperature, we plot $2\Delta(T)$ according to the temperature dependence in BCS theory. For the value $\Delta_0 = 7.8\pm0.7$~meV, the midpoint temperatures $T_\alpha$ of both U and M modes agree reasonably well with the relation $\hbar\omega_\alpha = 2\Delta(T_\alpha)$ (Fig.~\ref{fig2}(d)). The estimated gap gives the ratio $2\Delta_0/k_{\rm B}T_{\rm c} = 7.0$, which characterizes this compound as a strong coupling superconductor. According to the point-contact spectra on LaFeAsO$_{0.9}$F$_{0.1}$~\cite{gonnelli09}, two gaps have been reported; $\Delta_1 \approx$ 7.9~meV, and $\Delta_2 \approx$ 2.8~meV, both of which show strong deviation from the BCS-like temperature dependence. On the other hand, another experiment on SmFeAsO$_{0.85}$F$_{0.15}$~\cite{Chen2008} with $T_{\rm c}\sim 42$~K has reported a single gap around 7~meV that shows a BCS-like temperature dependence. The present result for $\Delta_0$ corresponds to the value $\Delta_1$ in \cite{gonnelli09}, but with BCS-like temperature dependence. Further study is necessary to reconcile these conflicting observations.

The crossing of $\hbar\omega$ and $2\Delta (T)$ is analogous to the optical threshold in the BCS superconductor~\cite{Mattis58}.  With finite damping of quasi-particles and/or presence of normal component of electrons, the optical conductivity becomes already finite~\cite{Li2008,Gorshunov2010,Herman17} for $\hbar\omega < 2\Delta_0$ in contrast with the ideal BCS case. In such a case, the midpoint of the increasing conductivity as a function of $\hbar\omega$ corresponds to the relation $\hbar\omega = 2\Delta_0$. The finite damping should also be present in the hydrogen excitations. Hence, the midpoint of $\hbar\Gamma_\alpha(T)$ seems to be a reasonable choice for extracting $2\Delta (T)$.

We proceed to consider the microscopic origin of the newly observed excitations. Firstly, to investigate stable positions of the interstitial hydrogen, we performed first-principles calculations using the Vienna ab initio simulation package, which is combined with a generalized gradient approximation using the Perdew--Burke--Ernzerhof exchange-correlation function~\cite{perdew}. The cut-off energy for the plane wave is set to 400~eV, and the $k$-point $3\times 3\times 3$ mesh is sampled in the first Brillouin zone. Interstitial hydrogens are placed in periodically repeated $2\times 2\times 1$ supercell corresponding to the interstitial hydrogen number 1/8 per unit cell. The interstitial hydrogen frequently induces displacements of surrounding atoms from their equilibrium positions, and forms a self-trapped state~\cite{Sundell04, Sundell07, Iwazaki10}. However, it has been discussed that thermal fluctuations recover the equivalent level for all interstitial sites, which enables the tunneling~\cite{Flynn70,Kagan74}. There is no consistent microscopic theory on how quantum tunneling occurs at low enough temperatures where thermal fluctuations are negligible. In view of this situation, the calculation was mainly performed using fixed positions of atoms. In the Supplemental Material, we discuss how the relaxation of surrounding atoms influences the result. To simulate the system with hydrogen implanted, we place a static point charge at location $\bm r$ in the supercell together with a neutralizing electron. Then, we compute the total energy $E (\bm r)$ per supercell, which is minimized at the optimal position $\bm r$ of the positive charge. We define the effective potential for a positive charge: $\Delta E(\bm r) = E (\bm r) - E_0-\mu$ where $E_0$ is the corresponding energy without the positive charge, and $\mu$ is the chemical potential. As a result, the dynamics of protons or deuterons are neglected in the present simulation.

\begin{figure}
\includegraphics[width=7.0cm]{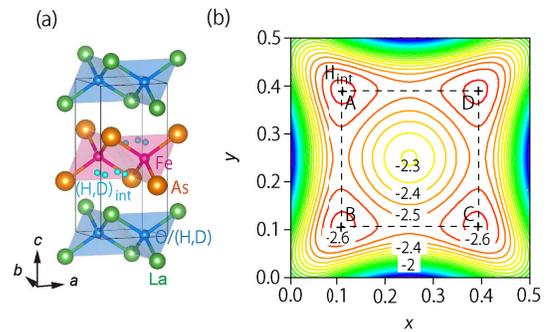}%
\caption{\label{fig3}
(a) Crystal structure of LaFeAsO$_{0.9}$D$_{0.1}$ with the interstitial site H$_{\rm int}$. (b) Contour map of $\Delta E$ in units of eV on the $xy$-plane at $z$ = 0.425. The four potential minima marked by the crosses are located at (0.11, 0.11, 0.425) and its equivalent positions. The fourfold rotational axis runs along the line (0.25, 0.25, $z$). The coordinates are described by the crystallographic units.
}
\end{figure}

According to structure analysis, LaFeAsO$_{0.9}$D$_{0.1}$ has a ZrCuSiAs-type structure with alternating stacks of conducting FeAs$_4$ and insulating (O,D)La$_4$ layers (Fig.~\ref{fig3}(a)). Figure~\ref{fig3}(b) illustrates the contour map of  $\Delta E(\bm r)$ on the $xy$-plane at $z$ = 0.425. There appear four potential minima in $\Delta E(\bm r)$ at $\bm r = (0.11, 0.11, 0.425)$ and its crystallographically equivalent positions related by the fourfold rotational symmetry, labeled H$_{\rm int}$ hereafter. The minima are located within the As$_5$La void with H$_{\rm int}$--As distance of 2.15~\AA. Bader charge analysis gives the charge of hydrogen at H$_{\rm int}$ to be 1.33, which implies the interstitial hydrogen lies in hydride~\cite{bader}.

\begin{figure}[h]
\includegraphics[width=7.5cm]{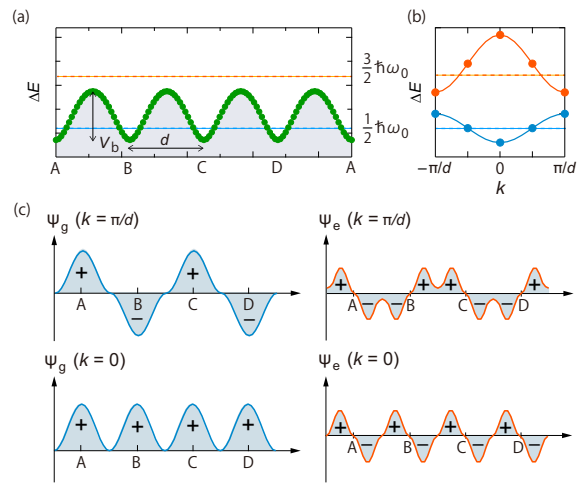}%
\caption{\label{fig4}
(a) The line profiles of $\Delta E$ through the minima A--B--C--D--A on $z$ = 0.425. The tunneling distance $d$ and the potential barrier $V_{\rm b}$ between the minima are indicated. The blue dashed line represents the ground state in the single harmonic potential at $\frac 12 \hbar\omega_0$, while the orange dashed line represents the first excited state at $\frac 32 \hbar\omega_0$. (b) The blue line shows the spectrum of the tunneling splitting, while the orange line shows that of the highly anharmonic phonons described as quantum rattling. (c) Schematic picture of hydrogen wave functions. $\Psi_{\rm g}$ stems from the ground state of the harmonic oscillator with $n=0$, while $\Psi_{\rm e}$ stems from $n=1$.  The wave functions at neighboring sites have either the same phase ($k=0$), or the opposite phase ($k=\pi/d$).
 }
 \end{figure}

Using the effective potential $\Delta E(\bm r)$, we now consider the dynamics of the interstitial hydrogen. Figure~\ref{fig4}(a) illustrates the line profile through the minima A--B--C--D--A, as indicated in Fig.~\ref{fig3}(b). This corresponds to the pathway of hydrogen migration. The profile features a periodic potential with tunneling distance $d$ = 1.13(1)~\AA\ and potential barrier $V_{\rm b}$ = 102~meV at the saddle point. In the harmonic approximation for sinusoidal potential, the angular frequency $\omega_0$ is given by $\omega_0^2$ = $2 \pi^2 V_{\rm b}/(md^2)$, where $m$ is the isotope mass. Next, we obtain $\hbar \omega_{\rm 0,H}$ = 81~meV for H and $\hbar \omega_{\rm 0,D}$ = 57~meV for D. The $V_{\rm b}$ is comparable to the 80~meV estimated for Nb(OH)$_x$ with $d$ = 1.17~\AA\ and $\hbar \omega_0$ = 107~meV~\cite{magerl83}.

The cross section $\sigma_{\rm inc}$ of incoherent scattering is much larger in H than D, with the ratio $\sigma_{\rm inc}({\rm H})/\sigma_{\rm inc}({\rm D}) \approx 40$. Considering the concentration ratio $n_{\rm H}/n_{\rm D} \approx 0.02$,  the scattering intensities from the two isotopes should be comparable. Hence, we assume that the features U and M come from the interstitial H and D, respectively. The ratio $\omega_{\rm U}/\omega_{\rm M} \approx 1.3$ is rather close to the square root of the atomic mass ratio $\sqrt{m_{\rm D}/m_{\rm H}} = \sqrt 2$. This might be regarded as the isotope effect for harmonic phonons. In fact, for H or D substituting oxygen, the Supplemental Material demonstrates the isotope effect with reasonable agreement between the calculation and the experimental observation in the range of 70--130~meV. However, since $\hbar\omega_{\rm U}$ and $\hbar\omega_{\rm M}$ are much smaller ($< 20$ meV), it is difficult to ascribe U and M to ordinary vibrations of hydrogen.

We now check the possibility of whether $\omega_{\rm U}$ and $\omega_{\rm M}$ may come from the tunneling motion of hydrogen. The magnitude of the tunneling splitting is sensitive to distance and mass of isotopes~\cite{fukai,wipf97,wipf84}. For example, the tunneling splittings are reported as 0.2~meV (H) and 0.02~meV (D) for Nb~\cite{magerl86,wipf84}, while  6.3~meV (H) and 1.6~meV (D) for $\alpha$-Mn~\cite{kolesnikov99}. The latter values are referred to as giant tunneling splittings, and are caused by the very short tunneling distance of 0.68~\AA. The mass of the isotopes affects the tunneling splittings $\omega_{\rm H,D}$ more strongly than the case for phonons, as seen in $\omega_{\rm H}/\omega_{\rm D}\approx 10$ for Nb~\cite{magerl86,wipf84} and $\omega_{\rm H}/\omega_{\rm D} \approx$ 4 for $\alpha$-Mn~\cite{kolesnikov99}. Thus, the present value $\omega_{\rm U}/\omega_{\rm M}\approx 1.3$ is not consistent with the identification of tunneling splitting.

We therefore propose that the excitations at $\hbar \omega_{\rm U}$ and $\hbar \omega_{\rm M}$ are associated with a new type of extremely anharmonic phonons, which can be appropriately called quantum rattling. In this process, the potential barrier is so low that the ordinary excited state, corresponding to the energy $\frac 32 \hbar\omega_0$ in the harmonic case, is actually extended along the pathway connecting the four potential minima, as illustrated in Fig.~\ref{fig4}(a). In four potential minima system, the eigenstates are characterized by the wave numbers $k=0, \pm \pi/(2d)$, and $\pm \pi/d$, as shown in Fig.~\ref{fig4}(b). In addition, the eigenstates are related to the superposition of localized phonons, which have energies $(n+1/2)\hbar\omega_0$ with $n=0,1,\ldots$ Figure~\ref{fig4}(c) illustrates the wave functions $\Psi_{\rm g}$ stemming from $n=0$, and $\Psi_{\rm e}$ stemming from $n=1$. The exact solution of the Schr\"{o}dinger equation is obtained in terms of the Mathieu functions. If we assign $\hbar \omega_{\rm M}=11.1$ meV as the deuterium excitation from $\Psi_{\rm g}(k=0)$ to $\Psi_{\rm e}(k=\pi/d)$, and assume the same potential barrier $V_{\rm b}$ for D and H, all other excitation energies are fixed theoretically. Hence, the corresponding protium excitation is expected at 16.6 meV, which compares favorably with the experimental value $\hbar \omega_{\rm U}$ = 14.5~meV. While, we have to take the value of $V_{\rm b} \approx 8$~meV, which is much smaller than the value 102~meV obtained from the first-principle calculation. We note that quantum fluctuations, which are neglected in this approximation, may lower the effective barrier substantially, as discussed in the literature~\cite{Sundell07}. Further study is required on the influence of fluctuation and self-trapping effects on the dynamics of hydrogen.

In conclusion, we found new excitations of hydrogen in the iron-based superconductor LaFeAsO$_{0.9}$D$_{0.1}$ using inelastic neutron scattering. We have ascribed the excitations to a novel quantum rattling, or a band motion of hydrogen, which arises only if the interstitial site has a number of potential minima larger than two. The new excitations of hydrogen broaden rapidly when twice the superconducting gap matches the energy of the hydrogen excitation. Assuming the BCS-like temperature dependence of $\Delta (T)$, we estimate the superconducting gap $\Delta_0 = 7.8$~meV which agrees with another probe result.

It is evident that the isotropic superconducting gap should be the simplest case for the present probe. Thus, the ion-based superconductors are much more suitable for the analysis than cuprates with nodes in the gap. Moreover, the matching of the superconducting gap with the hydrogen excitation is just fortunate in the iron-based superconductors with appropriate interstitial sites. The hydrogen spectroscopy has a unique feature to scan along the temperature axis with fixed energy, while most other spectroscopies scan along the energy axis with fixed temperature. With further refinement of both measurement and analysis, the hydrogen probe should provide more useful information for clarifying the superconducting gap. Concomitantly, the novel quantum motion of hydrogen reported in this paper deserves more detailed study in its own right.

\begin{acknowledgments}
We thank K.~Yamada and R.~Kajimoto for fruitful discussions. This work was supported by MEXT Elements Strategy Initiative to Form Core Research Center and JSPS KAKENHI (Nos. JP25287081 and 16K05434). The neutron and x-ray experiments at J-PARC/MLF and KEK/Photon Factory were performed under user programs No. 2013S0003, 2013A0002(U), 2014E0004, 2015E0002, and 2016S2-004. The DFT calculation has been performed under Large Scale Computer Simulation Program No.T12/13-02, 13/14-09, 14/15-13, 15/16-07, 16/17-18 (FY2012-2015) of KEK. The crystal structure is drawn by using the software VESTA~\cite{momma}.
\end{acknowledgments}

\bibliography{H-SCgap_prb_v15}

\providecommand{\noopsort}[1]{}\providecommand{\singleletter}[1]{#1}%
\begin{thebibliography}{41}%
\makeatletter
\providecommand \@ifxundefined [1]{%
 \@ifx{#1\undefined}
}%
\providecommand \@ifnum [1]{%
 \ifnum #1\expandafter \@firstoftwo
 \else \expandafter \@secondoftwo
 \fi
}%
\providecommand \@ifx [1]{%
 \ifx #1\expandafter \@firstoftwo
 \else \expandafter \@secondoftwo
 \fi
}%
\providecommand \natexlab [1]{#1}%
\providecommand \enquote  [1]{``#1''}%
\providecommand \bibnamefont  [1]{#1}%
\providecommand \bibfnamefont [1]{#1}%
\providecommand \citenamefont [1]{#1}%
\providecommand \href@noop [0]{\@secondoftwo}%
\providecommand \href [0]{\begingroup \@sanitize@url \@href}%
\providecommand \@href[1]{\@@startlink{#1}\@@href}%
\providecommand \@@href[1]{\endgroup#1\@@endlink}%
\providecommand \@sanitize@url [0]{\catcode `\\12\catcode `\$12\catcode
  `\&12\catcode `\#12\catcode `\^12\catcode `\_12\catcode `\%12\relax}%
\providecommand \@@startlink[1]{}%
\providecommand \@@endlink[0]{}%
\providecommand \url  [0]{\begingroup\@sanitize@url \@url }%
\providecommand \@url [1]{\endgroup\@href {#1}{\urlprefix }}%
\providecommand \urlprefix  [0]{URL }%
\providecommand \Eprint [0]{\href }%
\providecommand \doibase [0]{http://dx.doi.org/}%
\providecommand \selectlanguage [0]{\@gobble}%
\providecommand \bibinfo  [0]{\@secondoftwo}%
\providecommand \bibfield  [0]{\@secondoftwo}%
\providecommand \translation [1]{[#1]}%
\providecommand \BibitemOpen [0]{}%
\providecommand \bibitemStop [0]{}%
\providecommand \bibitemNoStop [0]{.\EOS\space}%
\providecommand \EOS [0]{\spacefactor3000\relax}%
\providecommand \BibitemShut  [1]{\csname bibitem#1\endcsname}%
\let\auto@bib@innerbib\@empty
\bibitem [{\citenamefont {Kamihara}\ \emph {et~al.}(2008)\citenamefont
  {Kamihara}, \citenamefont {Watanabe}, \citenamefont {Hirano},\ and\
  \citenamefont {Hosono}}]{kamihara}%
  \BibitemOpen
  \bibfield  {author} {\bibinfo {author} {\bibfnamefont {Y.}~\bibnamefont
  {Kamihara}}, \bibinfo {author} {\bibfnamefont {T.}~\bibnamefont {Watanabe}},
  \bibinfo {author} {\bibfnamefont {M.}~\bibnamefont {Hirano}}, \ and\ \bibinfo
  {author} {\bibfnamefont {H.}~\bibnamefont {Hosono}},\ }\href@noop {}
  {\bibfield  {journal} {\bibinfo  {journal} {J. Am. Chem. Soc.}\ }\textbf
  {\bibinfo {volume} {130}},\ \bibinfo {pages} {3296} (\bibinfo {year}
  {2008})}\BibitemShut {NoStop}%
\bibitem [{\citenamefont {Paglione}\ and\ \citenamefont
  {Greene}(2010)}]{paglione}%
  \BibitemOpen
  \bibfield  {author} {\bibinfo {author} {\bibfnamefont {J.}~\bibnamefont
  {Paglione}}\ and\ \bibinfo {author} {\bibfnamefont {R.~L.}\ \bibnamefont
  {Greene}},\ }\href@noop {} {\bibfield  {journal} {\bibinfo  {journal} {Nat.
  Phys.}\ }\textbf {\bibinfo {volume} {6}},\ \bibinfo {pages} {645} (\bibinfo
  {year} {2010})}\BibitemShut {NoStop}%
\bibitem [{\citenamefont {Stewart}(2011)}]{stewart}%
  \BibitemOpen
  \bibfield  {author} {\bibinfo {author} {\bibfnamefont {G.~R.}\ \bibnamefont
  {Stewart}},\ }\href@noop {} {\bibfield  {journal} {\bibinfo  {journal} {Rev.
  Mod. Phys.}\ }\textbf {\bibinfo {volume} {83}},\ \bibinfo {pages} {1589}
  (\bibinfo {year} {2011})}\BibitemShut {NoStop}%
\bibitem [{\citenamefont {Hosono}\ and\ \citenamefont {Kuroki}(2015)}]{hosono}%
  \BibitemOpen
  \bibfield  {author} {\bibinfo {author} {\bibfnamefont {H.}~\bibnamefont
  {Hosono}}\ and\ \bibinfo {author} {\bibfnamefont {K.}~\bibnamefont
  {Kuroki}},\ }\href@noop {} {\bibfield  {journal} {\bibinfo  {journal}
  {Physica C}\ }\textbf {\bibinfo {volume} {514}},\ \bibinfo {pages} {399}
  (\bibinfo {year} {2015})}\BibitemShut {NoStop}%
\bibitem [{\citenamefont {Ding}\ \emph {et~al.}(2008)\citenamefont {Ding},
  \citenamefont {Richard}, \citenamefont {Nakayama}, \citenamefont {Sugawara},
  \citenamefont {Arakane}, \citenamefont {Sekiba}, \citenamefont {Takayama},
  \citenamefont {Souma}, \citenamefont {Sato}, \citenamefont {Takahashi},
  \citenamefont {Wang}, \citenamefont {Dai}, \citenamefont {Fang},
  \citenamefont {Chen}, \citenamefont {Luo},\ and\ \citenamefont
  {Wang}}]{Ding2008}%
  \BibitemOpen
  \bibfield  {author} {\bibinfo {author} {\bibfnamefont {H.}~\bibnamefont
  {Ding}}, \bibinfo {author} {\bibfnamefont {P.}~\bibnamefont {Richard}},
  \bibinfo {author} {\bibfnamefont {K.}~\bibnamefont {Nakayama}}, \bibinfo
  {author} {\bibfnamefont {K.}~\bibnamefont {Sugawara}}, \bibinfo {author}
  {\bibfnamefont {T.}~\bibnamefont {Arakane}}, \bibinfo {author} {\bibfnamefont
  {Y.}~\bibnamefont {Sekiba}}, \bibinfo {author} {\bibfnamefont
  {A.}~\bibnamefont {Takayama}}, \bibinfo {author} {\bibfnamefont
  {S.}~\bibnamefont {Souma}}, \bibinfo {author} {\bibfnamefont
  {T.}~\bibnamefont {Sato}}, \bibinfo {author} {\bibfnamefont {T.}~\bibnamefont
  {Takahashi}}, \bibinfo {author} {\bibfnamefont {Z.}~\bibnamefont {Wang}},
  \bibinfo {author} {\bibfnamefont {X.}~\bibnamefont {Dai}}, \bibinfo {author}
  {\bibfnamefont {Z.}~\bibnamefont {Fang}}, \bibinfo {author} {\bibfnamefont
  {G.~F.}\ \bibnamefont {Chen}}, \bibinfo {author} {\bibfnamefont {J.~L.}\
  \bibnamefont {Luo}}, \ and\ \bibinfo {author} {\bibfnamefont {N.~L.}\
  \bibnamefont {Wang}},\ }\href {\doibase 10.1209/0295-5075/83/47001}
  {\bibfield  {journal} {\bibinfo  {journal} {Europhys. Lett.}\ }\textbf
  {\bibinfo {volume} {83}},\ \bibinfo {pages} {47001} (\bibinfo {year}
  {2008})}\BibitemShut {NoStop}%
\bibitem [{\citenamefont {Kondo}\ \emph {et~al.}(2008)\citenamefont {Kondo},
  \citenamefont {Santander-Syro}, \citenamefont {Copie}, \citenamefont {Liu},
  \citenamefont {Tillman}, \citenamefont {Mun}, \citenamefont {Schmalian},
  \citenamefont {Bud'ko}, \citenamefont {Tanatar}, \citenamefont {Canfield},\
  and\ \citenamefont {Kaminski}}]{Kondo2008}%
  \BibitemOpen
  \bibfield  {author} {\bibinfo {author} {\bibfnamefont {T.}~\bibnamefont
  {Kondo}}, \bibinfo {author} {\bibfnamefont {A.~F.}\ \bibnamefont
  {Santander-Syro}}, \bibinfo {author} {\bibfnamefont {O.}~\bibnamefont
  {Copie}}, \bibinfo {author} {\bibfnamefont {C.}~\bibnamefont {Liu}}, \bibinfo
  {author} {\bibfnamefont {M.~E.}\ \bibnamefont {Tillman}}, \bibinfo {author}
  {\bibfnamefont {E.~D.}\ \bibnamefont {Mun}}, \bibinfo {author} {\bibfnamefont
  {J.}~\bibnamefont {Schmalian}}, \bibinfo {author} {\bibfnamefont {S.~L.}\
  \bibnamefont {Bud'ko}}, \bibinfo {author} {\bibfnamefont {M.~A.}\
  \bibnamefont {Tanatar}}, \bibinfo {author} {\bibfnamefont {P.~C.}\
  \bibnamefont {Canfield}}, \ and\ \bibinfo {author} {\bibfnamefont
  {A.}~\bibnamefont {Kaminski}},\ }\href {\doibase
  10.1103/PhysRevLett.101.147003} {\bibfield  {journal} {\bibinfo  {journal}
  {Phys. Rev. Lett.}\ }\textbf {\bibinfo {volume} {101}},\ \bibinfo {pages}
  {147003} (\bibinfo {year} {2008})}\BibitemShut {NoStop}%
\bibitem [{\citenamefont {Hanaguri}\ \emph {et~al.}(2010)\citenamefont
  {Hanaguri}, \citenamefont {Niitaka}, \citenamefont {Kuroki},\ and\
  \citenamefont {Takagi}}]{Hanaguri2010}%
  \BibitemOpen
  \bibfield  {author} {\bibinfo {author} {\bibfnamefont {T.}~\bibnamefont
  {Hanaguri}}, \bibinfo {author} {\bibfnamefont {S.}~\bibnamefont {Niitaka}},
  \bibinfo {author} {\bibfnamefont {K.}~\bibnamefont {Kuroki}}, \ and\ \bibinfo
  {author} {\bibfnamefont {H.}~\bibnamefont {Takagi}},\ }\href {\doibase
  10.1126/science.1187399} {\bibfield  {journal} {\bibinfo  {journal}
  {Science}\ }\textbf {\bibinfo {volume} {328}},\ \bibinfo {pages} {474}
  (\bibinfo {year} {2010})}\BibitemShut {NoStop}%
\bibitem [{\citenamefont {Chen}\ \emph {et~al.}(2008)\citenamefont {Chen},
  \citenamefont {Tesanovic}, \citenamefont {Liu}, \citenamefont {Chen},\ and\
  \citenamefont {Chien}}]{Chen2008}%
  \BibitemOpen
  \bibfield  {author} {\bibinfo {author} {\bibfnamefont {T.~Y.}\ \bibnamefont
  {Chen}}, \bibinfo {author} {\bibfnamefont {Z.}~\bibnamefont {Tesanovic}},
  \bibinfo {author} {\bibfnamefont {R.~H.}\ \bibnamefont {Liu}}, \bibinfo
  {author} {\bibfnamefont {X.~H.}\ \bibnamefont {Chen}}, \ and\ \bibinfo
  {author} {\bibfnamefont {C.~L.}\ \bibnamefont {Chien}},\ }\href {\doibase
  10.1038/nature07081} {\bibfield  {journal} {\bibinfo  {journal} {Nature}\
  }\textbf {\bibinfo {volume} {453}},\ \bibinfo {pages} {1224} (\bibinfo {year}
  {2008})}\BibitemShut {NoStop}%
\bibitem [{\citenamefont {Gonnelli}\ \emph {et~al.}(2009)\citenamefont
  {Gonnelli}, \citenamefont {Tortello}, \citenamefont {Daghero}, \citenamefont
  {Ummarino}, \citenamefont {Stepanov},\ and\ \citenamefont
  {Kim}}]{gonnelli09}%
  \BibitemOpen
  \bibfield  {author} {\bibinfo {author} {\bibfnamefont {R.~S.}\ \bibnamefont
  {Gonnelli}}, \bibinfo {author} {\bibfnamefont {M.}~\bibnamefont {Tortello}},
  \bibinfo {author} {\bibfnamefont {D.}~\bibnamefont {Daghero}}, \bibinfo
  {author} {\bibfnamefont {G.~A.}\ \bibnamefont {Ummarino}}, \bibinfo {author}
  {\bibfnamefont {V.~A.}\ \bibnamefont {Stepanov}}, \ and\ \bibinfo {author}
  {\bibfnamefont {J.~S.}\ \bibnamefont {Kim}},\ }\href@noop {} {\bibfield
  {journal} {\bibinfo  {journal} {Cent. Eur. J. Phys.}\ }\textbf {\bibinfo
  {volume} {7}},\ \bibinfo {pages} {251} (\bibinfo {year} {2009})}\BibitemShut
  {NoStop}%
\bibitem [{\citenamefont {Li}\ \emph {et~al.}(2008)\citenamefont {Li},
  \citenamefont {Hu}, \citenamefont {Dong}, \citenamefont {Li}, \citenamefont
  {Zheng}, \citenamefont {Chen}, \citenamefont {Luo},\ and\ \citenamefont
  {Wang}}]{Li2008}%
  \BibitemOpen
  \bibfield  {author} {\bibinfo {author} {\bibfnamefont {G.}~\bibnamefont
  {Li}}, \bibinfo {author} {\bibfnamefont {W.~Z.}\ \bibnamefont {Hu}}, \bibinfo
  {author} {\bibfnamefont {J.}~\bibnamefont {Dong}}, \bibinfo {author}
  {\bibfnamefont {Z.}~\bibnamefont {Li}}, \bibinfo {author} {\bibfnamefont
  {P.}~\bibnamefont {Zheng}}, \bibinfo {author} {\bibfnamefont {G.~F.}\
  \bibnamefont {Chen}}, \bibinfo {author} {\bibfnamefont {J.~L.}\ \bibnamefont
  {Luo}}, \ and\ \bibinfo {author} {\bibfnamefont {N.~L.}\ \bibnamefont
  {Wang}},\ }\href {\doibase 10.1103/PhysRevLett.101.107004} {\bibfield
  {journal} {\bibinfo  {journal} {Phys. Rev. Lett.}\ }\textbf {\bibinfo
  {volume} {101}},\ \bibinfo {pages} {107004} (\bibinfo {year}
  {2008})}\BibitemShut {NoStop}%
\bibitem [{\citenamefont {Gorshunov}\ \emph {et~al.}(2010)\citenamefont
  {Gorshunov}, \citenamefont {Wu}, \citenamefont {Voronkov}, \citenamefont
  {Kallina}, \citenamefont {Iida}, \citenamefont {Haindl}, \citenamefont
  {Kurth}, \citenamefont {Schultz}, \citenamefont {Holzapfel},\ and\
  \citenamefont {Dressel}}]{Gorshunov2010}%
  \BibitemOpen
  \bibfield  {author} {\bibinfo {author} {\bibfnamefont {B.}~\bibnamefont
  {Gorshunov}}, \bibinfo {author} {\bibfnamefont {D.}~\bibnamefont {Wu}},
  \bibinfo {author} {\bibfnamefont {A.~A.}\ \bibnamefont {Voronkov}}, \bibinfo
  {author} {\bibfnamefont {P.}~\bibnamefont {Kallina}}, \bibinfo {author}
  {\bibfnamefont {K.}~\bibnamefont {Iida}}, \bibinfo {author} {\bibfnamefont
  {S.}~\bibnamefont {Haindl}}, \bibinfo {author} {\bibfnamefont
  {F.}~\bibnamefont {Kurth}}, \bibinfo {author} {\bibfnamefont
  {L.}~\bibnamefont {Schultz}}, \bibinfo {author} {\bibfnamefont
  {B.}~\bibnamefont {Holzapfel}}, \ and\ \bibinfo {author} {\bibfnamefont
  {M.}~\bibnamefont {Dressel}},\ }\href {\doibase 10.1103/PhysRevB.81.060509}
  {\bibfield  {journal} {\bibinfo  {journal} {Phys. Rev. B}\ }\textbf {\bibinfo
  {volume} {81}},\ \bibinfo {pages} {060509(R)} (\bibinfo {year}
  {2010})}\BibitemShut {NoStop}%
\bibitem [{\citenamefont {Fukai}(2003)}]{fukai}%
  \BibitemOpen
  \bibfield  {author} {\bibinfo {author} {\bibfnamefont {Y.}~\bibnamefont
  {Fukai}},\ }\href@noop {} {\emph {\bibinfo {title} {The Metal-Hydrogen
  System: Basic Bulk Properties}}}\ (\bibinfo  {publisher} {Springer},\
  \bibinfo {year} {2003})\BibitemShut {NoStop}%
\bibitem [{\citenamefont {Kehr}(1978)}]{kehr78}%
  \BibitemOpen
  \bibfield  {author} {\bibinfo {author} {\bibfnamefont {K.~W.}\ \bibnamefont
  {Kehr}},\ }in\ \href@noop {} {\emph {\bibinfo {booktitle} {Hydrogen in Metals
  I: Basic Properties}}},\ Vol.~\bibinfo {volume} {28},\ \bibinfo {editor}
  {edited by\ \bibinfo {editor} {\bibfnamefont {G.}~\bibnamefont {Alefeld}}\
  and\ \bibinfo {editor} {\bibfnamefont {J.}~\bibnamefont {Voelkl}}}\ (\bibinfo
   {publisher} {Springer, Berlin, Heidelberg},\ \bibinfo {year}
  {1978})\BibitemShut {NoStop}%
\bibitem [{\citenamefont {Wipf}(1997)}]{wipf97}%
  \BibitemOpen
  \bibinfo {editor} {\bibfnamefont {H.}~\bibnamefont {Wipf}},\ ed.,\ in\
  \href@noop {} {\emph {\bibinfo {booktitle} {Hydrogen in Metals III:
  Properties and Applications (Topics in Applied Physics)}}},\ Vol.~\bibinfo
  {volume} {73}\ (\bibinfo  {publisher} {Springer, Berlin, Heidelberg},\
  \bibinfo {year} {1997})\BibitemShut {NoStop}%
\bibitem [{\citenamefont {Kondo}(1986)}]{kondo86}%
  \BibitemOpen
  \bibfield  {author} {\bibinfo {author} {\bibfnamefont {J.}~\bibnamefont
  {Kondo}},\ }\href@noop {} {\bibfield  {journal} {\bibinfo  {journal} {Physica
  B}\ }\textbf {\bibinfo {volume} {141}},\ \bibinfo {pages} {305} (\bibinfo
  {year} {1986})}\BibitemShut {NoStop}%
\bibitem [{\citenamefont {Kagan}\ and\ \citenamefont
  {Prokof'ev}(1992)}]{kagan92}%
  \BibitemOpen
  \bibfield  {author} {\bibinfo {author} {\bibfnamefont {Y.}~\bibnamefont
  {Kagan}}\ and\ \bibinfo {author} {\bibfnamefont {N.~V.}\ \bibnamefont
  {Prokof'ev}},\ }in\ \href@noop {} {\emph {\bibinfo {booktitle} {Quantum
  tunnelling in condensed media}}},\ \bibinfo {editor} {edited by\ \bibinfo
  {editor} {\bibfnamefont {Y.}~\bibnamefont {Kagan}}\ and\ \bibinfo {editor}
  {\bibfnamefont {A.~J.}\ \bibnamefont {Leggett}}}\ (\bibinfo  {publisher}
  {North-Holland},\ \bibinfo {address} {Amsterdam $\cdot$ Lonodn $\cdot$ New
  York $\cdot$ Tokyo},\ \bibinfo {year} {1992})\ Chap.~\bibinfo {chapter} {2},
  pp.\ \bibinfo {pages} {37--144}\BibitemShut {NoStop}%
\bibitem [{\citenamefont {Iimura}\ \emph {et~al.}(2012)\citenamefont {Iimura},
  \citenamefont {Matuishi}, \citenamefont {Sato}, \citenamefont {Hanna},
  \citenamefont {Muraba}, \citenamefont {Kim}, \citenamefont {Kim},
  \citenamefont {Takata},\ and\ \citenamefont {Hosono}}]{iimura12}%
  \BibitemOpen
  \bibfield  {author} {\bibinfo {author} {\bibfnamefont {S.}~\bibnamefont
  {Iimura}}, \bibinfo {author} {\bibfnamefont {S.}~\bibnamefont {Matuishi}},
  \bibinfo {author} {\bibfnamefont {H.}~\bibnamefont {Sato}}, \bibinfo {author}
  {\bibfnamefont {T.}~\bibnamefont {Hanna}}, \bibinfo {author} {\bibfnamefont
  {Y.}~\bibnamefont {Muraba}}, \bibinfo {author} {\bibfnamefont {S.~W.}\
  \bibnamefont {Kim}}, \bibinfo {author} {\bibfnamefont {J.~E.}\ \bibnamefont
  {Kim}}, \bibinfo {author} {\bibfnamefont {M.}~\bibnamefont {Takata}}, \ and\
  \bibinfo {author} {\bibfnamefont {H.}~\bibnamefont {Hosono}},\ }\href@noop {}
  {\bibfield  {journal} {\bibinfo  {journal} {Nat. Commun.}\ }\textbf {\bibinfo
  {volume} {3}},\ \bibinfo {pages} {943} (\bibinfo {year} {2012})}\BibitemShut
  {NoStop}%
\bibitem [{\citenamefont {Matsuishi}\ \emph {et~al.}(2014)\citenamefont
  {Matsuishi}, \citenamefont {Maruyama}, \citenamefont {Iimura},\ and\
  \citenamefont {Hosono}}]{matsuishi}%
  \BibitemOpen
  \bibfield  {author} {\bibinfo {author} {\bibfnamefont {S.}~\bibnamefont
  {Matsuishi}}, \bibinfo {author} {\bibfnamefont {T.}~\bibnamefont {Maruyama}},
  \bibinfo {author} {\bibfnamefont {S.}~\bibnamefont {Iimura}}, \ and\ \bibinfo
  {author} {\bibfnamefont {H.}~\bibnamefont {Hosono}},\ }\href@noop {}
  {\bibfield  {journal} {\bibinfo  {journal} {Phys. Rev. B}\ }\textbf {\bibinfo
  {volume} {89}},\ \bibinfo {pages} {094510} (\bibinfo {year}
  {2014})}\BibitemShut {NoStop}%
\bibitem [{\citenamefont {Hiraishi}\ \emph {et~al.}(2014)\citenamefont
  {Hiraishi}, \citenamefont {Iimura}, \citenamefont {Kojima}, \citenamefont
  {Yamaura}, \citenamefont {Hiraka}, \citenamefont {Ikeda}, \citenamefont
  {Miao}, \citenamefont {Ishikawa}, \citenamefont {Torii}, \citenamefont
  {Miyazaki}, \citenamefont {Yamauchi}, \citenamefont {Koda}, \citenamefont
  {Ishii}, \citenamefont {Yoshida}, \citenamefont {Mizuki}, \citenamefont
  {Kadono}, \citenamefont {Kumai}, \citenamefont {Kamiyama}, \citenamefont
  {Otomo}, \citenamefont {Murakami}, \citenamefont {Matuishi},\ and\
  \citenamefont {Hosono}}]{hiraishi14}%
  \BibitemOpen
  \bibfield  {author} {\bibinfo {author} {\bibfnamefont {M.}~\bibnamefont
  {Hiraishi}}, \bibinfo {author} {\bibfnamefont {S.}~\bibnamefont {Iimura}},
  \bibinfo {author} {\bibfnamefont {K.~M.}\ \bibnamefont {Kojima}}, \bibinfo
  {author} {\bibfnamefont {J.}~\bibnamefont {Yamaura}}, \bibinfo {author}
  {\bibfnamefont {H.}~\bibnamefont {Hiraka}}, \bibinfo {author} {\bibfnamefont
  {K.}~\bibnamefont {Ikeda}}, \bibinfo {author} {\bibfnamefont
  {P.}~\bibnamefont {Miao}}, \bibinfo {author} {\bibfnamefont {Y.}~\bibnamefont
  {Ishikawa}}, \bibinfo {author} {\bibfnamefont {S.}~\bibnamefont {Torii}},
  \bibinfo {author} {\bibfnamefont {M.}~\bibnamefont {Miyazaki}}, \bibinfo
  {author} {\bibfnamefont {I.}~\bibnamefont {Yamauchi}}, \bibinfo {author}
  {\bibfnamefont {A.}~\bibnamefont {Koda}}, \bibinfo {author} {\bibfnamefont
  {K.}~\bibnamefont {Ishii}}, \bibinfo {author} {\bibfnamefont
  {M.}~\bibnamefont {Yoshida}}, \bibinfo {author} {\bibfnamefont
  {J.}~\bibnamefont {Mizuki}}, \bibinfo {author} {\bibfnamefont
  {R.}~\bibnamefont {Kadono}}, \bibinfo {author} {\bibfnamefont
  {R.}~\bibnamefont {Kumai}}, \bibinfo {author} {\bibfnamefont
  {T.}~\bibnamefont {Kamiyama}}, \bibinfo {author} {\bibfnamefont
  {T.}~\bibnamefont {Otomo}}, \bibinfo {author} {\bibfnamefont
  {Y.}~\bibnamefont {Murakami}}, \bibinfo {author} {\bibfnamefont
  {S.}~\bibnamefont {Matuishi}}, \ and\ \bibinfo {author} {\bibfnamefont
  {H.}~\bibnamefont {Hosono}},\ }\href@noop {} {\bibfield  {journal} {\bibinfo
  {journal} {Nat. Phys.}\ }\textbf {\bibinfo {volume} {10}},\ \bibinfo {pages}
  {300} (\bibinfo {year} {2014})}\BibitemShut {NoStop}%
\bibitem [{\citenamefont {Iimura}\ \emph {et~al.}(2013)\citenamefont {Iimura},
  \citenamefont {Matsuishi}, \citenamefont {Miyakawa}, \citenamefont
  {Taniguchi}, \citenamefont {Suzuki}, \citenamefont {Usui}, \citenamefont
  {Kuroki}, \citenamefont {Kajimoto}, \citenamefont {Nakamura}, \citenamefont
  {Inamura}, \citenamefont {Ikeuchi}, \citenamefont {Ji},\ and\ \citenamefont
  {Hosono}}]{iimura13}%
  \BibitemOpen
  \bibfield  {author} {\bibinfo {author} {\bibfnamefont {S.}~\bibnamefont
  {Iimura}}, \bibinfo {author} {\bibfnamefont {S.}~\bibnamefont {Matsuishi}},
  \bibinfo {author} {\bibfnamefont {M.}~\bibnamefont {Miyakawa}}, \bibinfo
  {author} {\bibfnamefont {T.}~\bibnamefont {Taniguchi}}, \bibinfo {author}
  {\bibfnamefont {K.}~\bibnamefont {Suzuki}}, \bibinfo {author} {\bibfnamefont
  {H.}~\bibnamefont {Usui}}, \bibinfo {author} {\bibfnamefont {K.}~\bibnamefont
  {Kuroki}}, \bibinfo {author} {\bibfnamefont {R.}~\bibnamefont {Kajimoto}},
  \bibinfo {author} {\bibfnamefont {M.}~\bibnamefont {Nakamura}}, \bibinfo
  {author} {\bibfnamefont {Y.}~\bibnamefont {Inamura}}, \bibinfo {author}
  {\bibfnamefont {K.}~\bibnamefont {Ikeuchi}}, \bibinfo {author} {\bibfnamefont
  {S.}~\bibnamefont {Ji}}, \ and\ \bibinfo {author} {\bibfnamefont
  {H.}~\bibnamefont {Hosono}},\ }\href@noop {} {\bibfield  {journal} {\bibinfo
  {journal} {Phys. Rev. B}\ }\textbf {\bibinfo {volume} {88}},\ \bibinfo
  {pages} {060501(R)} (\bibinfo {year} {2013})}\BibitemShut {NoStop}%
\bibitem [{\citenamefont {Kajimoto}\ \emph {et~al.}(2011)\citenamefont
  {Kajimoto}, \citenamefont {Nakamura}, \citenamefont {Inamura}, \citenamefont
  {Mizuno}, \citenamefont {Nakajima}, \citenamefont {Ohira-Kawamura},
  \citenamefont {Yokoo}, \citenamefont {Nakatani}, \citenamefont {Maruyama},
  \citenamefont {Soyama}, \citenamefont {Shibata}, \citenamefont {Suzuya},
  \citenamefont {Sato}, \citenamefont {Aizawa}, \citenamefont {Arai},
  \citenamefont {Wakimoto}, \citenamefont {Ishikado}, \citenamefont {Shamoto},
  \citenamefont {Fujita}, \citenamefont {Hiraka}, \citenamefont {Ohoyama},
  \citenamefont {Yamada},\ and\ \citenamefont {Lee}}]{4seasons}%
  \BibitemOpen
  \bibfield  {author} {\bibinfo {author} {\bibfnamefont {R.}~\bibnamefont
  {Kajimoto}}, \bibinfo {author} {\bibfnamefont {M.}~\bibnamefont {Nakamura}},
  \bibinfo {author} {\bibfnamefont {Y.}~\bibnamefont {Inamura}}, \bibinfo
  {author} {\bibfnamefont {F.}~\bibnamefont {Mizuno}}, \bibinfo {author}
  {\bibfnamefont {K.}~\bibnamefont {Nakajima}}, \bibinfo {author}
  {\bibfnamefont {S.}~\bibnamefont {Ohira-Kawamura}}, \bibinfo {author}
  {\bibfnamefont {T.}~\bibnamefont {Yokoo}}, \bibinfo {author} {\bibfnamefont
  {T.}~\bibnamefont {Nakatani}}, \bibinfo {author} {\bibfnamefont
  {R.}~\bibnamefont {Maruyama}}, \bibinfo {author} {\bibfnamefont
  {K.}~\bibnamefont {Soyama}}, \bibinfo {author} {\bibfnamefont
  {K.}~\bibnamefont {Shibata}}, \bibinfo {author} {\bibfnamefont
  {K.}~\bibnamefont {Suzuya}}, \bibinfo {author} {\bibfnamefont
  {S.}~\bibnamefont {Sato}}, \bibinfo {author} {\bibfnamefont {K.}~\bibnamefont
  {Aizawa}}, \bibinfo {author} {\bibfnamefont {M.}~\bibnamefont {Arai}},
  \bibinfo {author} {\bibfnamefont {S.}~\bibnamefont {Wakimoto}}, \bibinfo
  {author} {\bibfnamefont {M.}~\bibnamefont {Ishikado}}, \bibinfo {author}
  {\bibfnamefont {S.}~\bibnamefont {Shamoto}}, \bibinfo {author} {\bibfnamefont
  {M.}~\bibnamefont {Fujita}}, \bibinfo {author} {\bibfnamefont
  {H.}~\bibnamefont {Hiraka}}, \bibinfo {author} {\bibfnamefont
  {K.}~\bibnamefont {Ohoyama}}, \bibinfo {author} {\bibfnamefont
  {K.}~\bibnamefont {Yamada}}, \ and\ \bibinfo {author} {\bibfnamefont {C.~H.}\
  \bibnamefont {Lee}},\ }\href@noop {} {\bibfield  {journal} {\bibinfo
  {journal} {J. Phys. Soc. Jpn.}\ }\textbf {\bibinfo {volume} {80}},\ \bibinfo
  {pages} {SB025} (\bibinfo {year} {2011})}\BibitemShut {NoStop}%
\bibitem [{\citenamefont {Nakamura}\ \emph {et~al.}(2009)\citenamefont
  {Nakamura}, \citenamefont {Kajimoto}, \citenamefont {Inamura}, \citenamefont
  {Mizuno}, \citenamefont {Fujita}, \citenamefont {Yokoo},\ and\ \citenamefont
  {Arai}}]{nakamura09}%
  \BibitemOpen
  \bibfield  {author} {\bibinfo {author} {\bibfnamefont {M.}~\bibnamefont
  {Nakamura}}, \bibinfo {author} {\bibfnamefont {R.}~\bibnamefont {Kajimoto}},
  \bibinfo {author} {\bibfnamefont {Y.}~\bibnamefont {Inamura}}, \bibinfo
  {author} {\bibfnamefont {F.}~\bibnamefont {Mizuno}}, \bibinfo {author}
  {\bibfnamefont {M.}~\bibnamefont {Fujita}}, \bibinfo {author} {\bibfnamefont
  {T.}~\bibnamefont {Yokoo}}, \ and\ \bibinfo {author} {\bibfnamefont
  {M.}~\bibnamefont {Arai}},\ }\href@noop {} {\bibfield  {journal} {\bibinfo
  {journal} {J. Phys. Soc. Jpn.}\ }\textbf {\bibinfo {volume} {78}},\ \bibinfo
  {pages} {093002} (\bibinfo {year} {2009})}\BibitemShut {NoStop}%
\bibitem [{\citenamefont {Inamura}\ \emph {et~al.}(2013)\citenamefont
  {Inamura}, \citenamefont {Nakatani}, \citenamefont {Suzuki},\ and\
  \citenamefont {Otomo}}]{inamura}%
  \BibitemOpen
  \bibfield  {author} {\bibinfo {author} {\bibfnamefont {Y.}~\bibnamefont
  {Inamura}}, \bibinfo {author} {\bibfnamefont {T.}~\bibnamefont {Nakatani}},
  \bibinfo {author} {\bibfnamefont {J.}~\bibnamefont {Suzuki}}, \ and\ \bibinfo
  {author} {\bibfnamefont {T.}~\bibnamefont {Otomo}},\ }\href@noop {}
  {\bibfield  {journal} {\bibinfo  {journal} {J. Phys. Soc. Jpn.}\ }\textbf
  {\bibinfo {volume} {82}},\ \bibinfo {pages} {SA031} (\bibinfo {year}
  {2013})}\BibitemShut {NoStop}%
\bibitem [{\citenamefont {Christianson}\ \emph {et~al.}(2008)\citenamefont
  {Christianson}, \citenamefont {Lumsden}, \citenamefont {Delaire},
  \citenamefont {Stone}, \citenamefont {Abernathy}, \citenamefont {McGuire},
  \citenamefont {Sefat}, \citenamefont {Jin}, \citenamefont {Sales},
  \citenamefont {Mandrus}, \citenamefont {Mun}, \citenamefont {Canfield},
  \citenamefont {Lin}, \citenamefont {Lucas}, \citenamefont {Kresch},
  \citenamefont {Keith}, \citenamefont {Fultz}, \citenamefont {Goremychkin},\
  and\ \citenamefont {McQueeney}}]{christianson08}%
  \BibitemOpen
  \bibfield  {author} {\bibinfo {author} {\bibfnamefont {A.~D.}\ \bibnamefont
  {Christianson}}, \bibinfo {author} {\bibfnamefont {M.~D.}\ \bibnamefont
  {Lumsden}}, \bibinfo {author} {\bibfnamefont {O.}~\bibnamefont {Delaire}},
  \bibinfo {author} {\bibfnamefont {M.~B.}\ \bibnamefont {Stone}}, \bibinfo
  {author} {\bibfnamefont {D.~L.}\ \bibnamefont {Abernathy}}, \bibinfo {author}
  {\bibfnamefont {M.~A.}\ \bibnamefont {McGuire}}, \bibinfo {author}
  {\bibfnamefont {A.~S.}\ \bibnamefont {Sefat}}, \bibinfo {author}
  {\bibfnamefont {R.}~\bibnamefont {Jin}}, \bibinfo {author} {\bibfnamefont
  {B.~C.}\ \bibnamefont {Sales}}, \bibinfo {author} {\bibfnamefont
  {D.}~\bibnamefont {Mandrus}}, \bibinfo {author} {\bibfnamefont {E.~D.}\
  \bibnamefont {Mun}}, \bibinfo {author} {\bibfnamefont {P.~C.}\ \bibnamefont
  {Canfield}}, \bibinfo {author} {\bibfnamefont {J.~Y.~Y.}\ \bibnamefont
  {Lin}}, \bibinfo {author} {\bibfnamefont {M.}~\bibnamefont {Lucas}}, \bibinfo
  {author} {\bibfnamefont {M.}~\bibnamefont {Kresch}}, \bibinfo {author}
  {\bibfnamefont {J.~B.}\ \bibnamefont {Keith}}, \bibinfo {author}
  {\bibfnamefont {B.}~\bibnamefont {Fultz}}, \bibinfo {author} {\bibfnamefont
  {E.~A.}\ \bibnamefont {Goremychkin}}, \ and\ \bibinfo {author} {\bibfnamefont
  {R.~J.}\ \bibnamefont {McQueeney}},\ }\href@noop {} {\bibfield  {journal}
  {\bibinfo  {journal} {Phys. Rev. Lett.}\ }\textbf {\bibinfo {volume} {101}},\
  \bibinfo {pages} {157004} (\bibinfo {year} {2008})}\BibitemShut {NoStop}%
\bibitem [{\citenamefont {Shamoto}\ \emph {et~al.}(2010)\citenamefont
  {Shamoto}, \citenamefont {Ishikado}, \citenamefont {Christianson},
  \citenamefont {Lumsden}, \citenamefont {Wakimoto}, \citenamefont {Kodama},
  \citenamefont {Iyo},\ and\ \citenamefont {Arai}}]{shamoto10}%
  \BibitemOpen
  \bibfield  {author} {\bibinfo {author} {\bibfnamefont {S.~I.}\ \bibnamefont
  {Shamoto}}, \bibinfo {author} {\bibfnamefont {M.}~\bibnamefont {Ishikado}},
  \bibinfo {author} {\bibfnamefont {A.~D.}\ \bibnamefont {Christianson}},
  \bibinfo {author} {\bibfnamefont {M.~D.}\ \bibnamefont {Lumsden}}, \bibinfo
  {author} {\bibfnamefont {S.}~\bibnamefont {Wakimoto}}, \bibinfo {author}
  {\bibfnamefont {K.}~\bibnamefont {Kodama}}, \bibinfo {author} {\bibfnamefont
  {A.}~\bibnamefont {Iyo}}, \ and\ \bibinfo {author} {\bibfnamefont
  {M.}~\bibnamefont {Arai}},\ }\href@noop {} {\bibfield  {journal} {\bibinfo
  {journal} {Phys. Rev. B}\ }\textbf {\bibinfo {volume} {82}},\ \bibinfo
  {pages} {172508} (\bibinfo {year} {2010})}\BibitemShut {NoStop}%
\bibitem [{\citenamefont {Wakimoto}\ \emph {et~al.}(2010)\citenamefont
  {Wakimoto}, \citenamefont {Kodama}, \citenamefont {Ishikado}, \citenamefont
  {Matsuda}, \citenamefont {Kajimoto}, \citenamefont {Arai}, \citenamefont
  {Kakurai}, \citenamefont {Esaka}, \citenamefont {Iyo}, \citenamefont {Kito},
  \citenamefont {Eisaki},\ and\ \citenamefont {Shamoto}}]{wakimoto10}%
  \BibitemOpen
  \bibfield  {author} {\bibinfo {author} {\bibfnamefont {S.}~\bibnamefont
  {Wakimoto}}, \bibinfo {author} {\bibfnamefont {K.}~\bibnamefont {Kodama}},
  \bibinfo {author} {\bibfnamefont {M.}~\bibnamefont {Ishikado}}, \bibinfo
  {author} {\bibfnamefont {M.}~\bibnamefont {Matsuda}}, \bibinfo {author}
  {\bibfnamefont {R.}~\bibnamefont {Kajimoto}}, \bibinfo {author}
  {\bibfnamefont {M.}~\bibnamefont {Arai}}, \bibinfo {author} {\bibfnamefont
  {K.}~\bibnamefont {Kakurai}}, \bibinfo {author} {\bibfnamefont
  {F.}~\bibnamefont {Esaka}}, \bibinfo {author} {\bibfnamefont
  {A.}~\bibnamefont {Iyo}}, \bibinfo {author} {\bibfnamefont {H.}~\bibnamefont
  {Kito}}, \bibinfo {author} {\bibfnamefont {H.}~\bibnamefont {Eisaki}}, \ and\
  \bibinfo {author} {\bibfnamefont {S.}~\bibnamefont {Shamoto}},\ }\href@noop
  {} {\bibfield  {journal} {\bibinfo  {journal} {J. Phys. Soc. Jpn.}\ }\textbf
  {\bibinfo {volume} {79}},\ \bibinfo {pages} {074715} (\bibinfo {year}
  {2010})}\BibitemShut {NoStop}%
\bibitem [{\citenamefont {Magerl}\ \emph {et~al.}(1986)\citenamefont {Magerl},
  \citenamefont {Dianoux}, \citenamefont {Wipf}, \citenamefont {Neumaier},\
  and\ \citenamefont {Anderson}}]{magerl86}%
  \BibitemOpen
  \bibfield  {author} {\bibinfo {author} {\bibfnamefont {A.}~\bibnamefont
  {Magerl}}, \bibinfo {author} {\bibfnamefont {A.~J.}\ \bibnamefont {Dianoux}},
  \bibinfo {author} {\bibfnamefont {H.}~\bibnamefont {Wipf}}, \bibinfo {author}
  {\bibfnamefont {K.}~\bibnamefont {Neumaier}}, \ and\ \bibinfo {author}
  {\bibfnamefont {I.~S.}\ \bibnamefont {Anderson}},\ }\href@noop {} {\bibfield
  {journal} {\bibinfo  {journal} {Phys. Rev. Lett.}\ }\textbf {\bibinfo
  {volume} {56}},\ \bibinfo {pages} {159} (\bibinfo {year} {1986})}\BibitemShut
  {NoStop}%
\bibitem [{\citenamefont {Black}\ and\ \citenamefont {Fulde}(1979)}]{Black79}%
  \BibitemOpen
  \bibfield  {author} {\bibinfo {author} {\bibfnamefont {J.~L.}\ \bibnamefont
  {Black}}\ and\ \bibinfo {author} {\bibfnamefont {P.}~\bibnamefont {Fulde}},\
  }\href@noop {} {\bibfield  {journal} {\bibinfo  {journal} {Phys. Rev. Lett.}\
  }\textbf {\bibinfo {volume} {43}},\ \bibinfo {pages} {453} (\bibinfo {year}
  {1979})}\BibitemShut {NoStop}%
\bibitem [{\citenamefont {Magerl}\ \emph {et~al.}(1983)\citenamefont {Magerl},
  \citenamefont {Rush}, \citenamefont {Rowe}, \citenamefont {Richter},\ and\
  \citenamefont {Wipf}}]{magerl83}%
  \BibitemOpen
  \bibfield  {author} {\bibinfo {author} {\bibfnamefont {A.}~\bibnamefont
  {Magerl}}, \bibinfo {author} {\bibfnamefont {J.~J.}\ \bibnamefont {Rush}},
  \bibinfo {author} {\bibfnamefont {J.~M.}\ \bibnamefont {Rowe}}, \bibinfo
  {author} {\bibfnamefont {D.}~\bibnamefont {Richter}}, \ and\ \bibinfo
  {author} {\bibfnamefont {H.}~\bibnamefont {Wipf}},\ }\href@noop {} {\bibfield
   {journal} {\bibinfo  {journal} {Phys. Rev. B}\ }\textbf {\bibinfo {volume}
  {27}},\ \bibinfo {pages} {927} (\bibinfo {year} {1983})}\BibitemShut
  {NoStop}%
\bibitem [{\citenamefont {Mattis}\ and\ \citenamefont
  {Bardeen}(1958)}]{Mattis58}%
  \BibitemOpen
  \bibfield  {author} {\bibinfo {author} {\bibfnamefont {D.~C.}\ \bibnamefont
  {Mattis}}\ and\ \bibinfo {author} {\bibfnamefont {J.}~\bibnamefont
  {Bardeen}},\ }\href@noop {} {\bibfield  {journal} {\bibinfo  {journal} {Phys.
  Rev.}\ }\textbf {\bibinfo {volume} {111}},\ \bibinfo {pages} {412} (\bibinfo
  {year} {1958})}\BibitemShut {NoStop}%
\bibitem [{\citenamefont {Herman}\ and\ \citenamefont
  {Hlubina}(2017)}]{Herman17}%
  \BibitemOpen
  \bibfield  {author} {\bibinfo {author} {\bibfnamefont {F.}~\bibnamefont
  {Herman}}\ and\ \bibinfo {author} {\bibfnamefont {R.}~\bibnamefont
  {Hlubina}},\ }\href@noop {} {\bibfield  {journal} {\bibinfo  {journal} {Phys.
  Rev. B}\ }\textbf {\bibinfo {volume} {96}},\ \bibinfo {pages} {014509}
  (\bibinfo {year} {2017})}\BibitemShut {NoStop}%
\bibitem [{\citenamefont {Perdew}\ \emph {et~al.}(1996)\citenamefont {Perdew},
  \citenamefont {Burke},\ and\ \citenamefont {Ernzerhof}}]{perdew}%
  \BibitemOpen
  \bibfield  {author} {\bibinfo {author} {\bibfnamefont {J.~P.}\ \bibnamefont
  {Perdew}}, \bibinfo {author} {\bibfnamefont {K.}~\bibnamefont {Burke}}, \
  and\ \bibinfo {author} {\bibfnamefont {M.}~\bibnamefont {Ernzerhof}},\
  }\href@noop {} {\bibfield  {journal} {\bibinfo  {journal} {Phys. Rev. Lett.}\
  }\textbf {\bibinfo {volume} {77}},\ \bibinfo {pages} {3865} (\bibinfo {year}
  {1996})}\BibitemShut {NoStop}%
\bibitem [{\citenamefont {Sundell}\ and\ \citenamefont
  {Wahnstr\"om}(2004)}]{Sundell04}%
  \BibitemOpen
  \bibfield  {author} {\bibinfo {author} {\bibfnamefont {P.~G.}\ \bibnamefont
  {Sundell}}\ and\ \bibinfo {author} {\bibfnamefont {G.}~\bibnamefont
  {Wahnstr\"om}},\ }\href@noop {} {\bibfield  {journal} {\bibinfo  {journal}
  {Phys. Rev. B}\ }\textbf {\bibinfo {volume} {70}},\ \bibinfo {pages} {224301}
  (\bibinfo {year} {2004})}\BibitemShut {NoStop}%
\bibitem [{\citenamefont {Sundell}\ \emph {et~al.}(2007)\citenamefont
  {Sundell}, \citenamefont {Bj\"orketun},\ and\ \citenamefont
  {Wahnstr\"om}}]{Sundell07}%
  \BibitemOpen
  \bibfield  {author} {\bibinfo {author} {\bibfnamefont {P.~G.}\ \bibnamefont
  {Sundell}}, \bibinfo {author} {\bibfnamefont {M.~E.}\ \bibnamefont
  {Bj\"orketun}}, \ and\ \bibinfo {author} {\bibfnamefont {G.}~\bibnamefont
  {Wahnstr\"om}},\ }\href@noop {} {\bibfield  {journal} {\bibinfo  {journal}
  {Phys. Rev. B}\ }\textbf {\bibinfo {volume} {76}},\ \bibinfo {pages} {094301}
  (\bibinfo {year} {2007})}\BibitemShut {NoStop}%
\bibitem [{\citenamefont {Iwazaki}\ \emph {et~al.}(2010)\citenamefont
  {Iwazaki}, \citenamefont {Suzuki},\ and\ \citenamefont
  {Tsuneyuki}}]{Iwazaki10}%
  \BibitemOpen
  \bibfield  {author} {\bibinfo {author} {\bibfnamefont {Y.}~\bibnamefont
  {Iwazaki}}, \bibinfo {author} {\bibfnamefont {T.}~\bibnamefont {Suzuki}}, \
  and\ \bibinfo {author} {\bibfnamefont {S.}~\bibnamefont {Tsuneyuki}},\
  }\href@noop {} {\bibfield  {journal} {\bibinfo  {journal} {J. Appl. Phys.}\
  }\textbf {\bibinfo {volume} {108}},\ \bibinfo {pages} {083705} (\bibinfo
  {year} {2010})}\BibitemShut {NoStop}%
\bibitem [{\citenamefont {Flynn}\ and\ \citenamefont
  {Stoneham}(1970)}]{Flynn70}%
  \BibitemOpen
  \bibfield  {author} {\bibinfo {author} {\bibfnamefont {C.~P.}\ \bibnamefont
  {Flynn}}\ and\ \bibinfo {author} {\bibfnamefont {A.~M.}\ \bibnamefont
  {Stoneham}},\ }\href {\doibase 10.1103/PhysRevB.1.3966} {\bibfield  {journal}
  {\bibinfo  {journal} {Phys. Rev. B}\ }\textbf {\bibinfo {volume} {1}},\
  \bibinfo {pages} {3966} (\bibinfo {year} {1970})}\BibitemShut {NoStop}%
\bibitem [{\citenamefont {Kagan}\ and\ \citenamefont
  {Klinger}(1974)}]{Kagan74}%
  \BibitemOpen
  \bibfield  {author} {\bibinfo {author} {\bibfnamefont {Y.}~\bibnamefont
  {Kagan}}\ and\ \bibinfo {author} {\bibfnamefont {M.~I.}\ \bibnamefont
  {Klinger}},\ }\href@noop {} {\bibfield  {journal} {\bibinfo  {journal} {J.
  Phys. C}\ }\textbf {\bibinfo {volume} {7}},\ \bibinfo {pages} {2791}
  (\bibinfo {year} {1974})}\BibitemShut {NoStop}%
\bibitem [{\citenamefont {Tang}\ \emph {et~al.}(2009)\citenamefont {Tang},
  \citenamefont {Sanville},\ and\ \citenamefont {Henkelman}}]{bader}%
  \BibitemOpen
  \bibfield  {author} {\bibinfo {author} {\bibfnamefont {W.}~\bibnamefont
  {Tang}}, \bibinfo {author} {\bibfnamefont {E.}~\bibnamefont {Sanville}}, \
  and\ \bibinfo {author} {\bibfnamefont {G.}~\bibnamefont {Henkelman}},\
  }\href@noop {} {\bibfield  {journal} {\bibinfo  {journal} {J. Phys.: Condens.
  Matter}\ }\textbf {\bibinfo {volume} {21}},\ \bibinfo {pages} {084204}
  (\bibinfo {year} {2009})}\BibitemShut {NoStop}%
\bibitem [{\citenamefont {Wipf}\ and\ \citenamefont {Neumaier}(1984)}]{wipf84}%
  \BibitemOpen
  \bibfield  {author} {\bibinfo {author} {\bibfnamefont {H.}~\bibnamefont
  {Wipf}}\ and\ \bibinfo {author} {\bibfnamefont {K.}~\bibnamefont
  {Neumaier}},\ }\href@noop {} {\bibfield  {journal} {\bibinfo  {journal}
  {Phys. Rev. Lett.}\ }\textbf {\bibinfo {volume} {52}},\ \bibinfo {pages}
  {1308} (\bibinfo {year} {1984})}\BibitemShut {NoStop}%
\bibitem [{\citenamefont {Kolesnikov}\ \emph {et~al.}(1999)\citenamefont
  {Kolesnikov}, \citenamefont {Antonov}, \citenamefont {Bennington},
  \citenamefont {Dorner}, \citenamefont {Fedotov}, \citenamefont {Grosse},
  \citenamefont {Li}, \citenamefont {Parker},\ and\ \citenamefont
  {Wagner}}]{kolesnikov99}%
  \BibitemOpen
  \bibfield  {author} {\bibinfo {author} {\bibfnamefont {A.~I.}\ \bibnamefont
  {Kolesnikov}}, \bibinfo {author} {\bibfnamefont {V.~E.}\ \bibnamefont
  {Antonov}}, \bibinfo {author} {\bibfnamefont {S.~M.}\ \bibnamefont
  {Bennington}}, \bibinfo {author} {\bibfnamefont {B.}~\bibnamefont {Dorner}},
  \bibinfo {author} {\bibfnamefont {V.~K.}\ \bibnamefont {Fedotov}}, \bibinfo
  {author} {\bibfnamefont {G.}~\bibnamefont {Grosse}}, \bibinfo {author}
  {\bibfnamefont {J.~C.}\ \bibnamefont {Li}}, \bibinfo {author} {\bibfnamefont
  {S.~F.}\ \bibnamefont {Parker}}, \ and\ \bibinfo {author} {\bibfnamefont
  {F.~E.}\ \bibnamefont {Wagner}},\ }\href@noop {} {\bibfield  {journal}
  {\bibinfo  {journal} {Physica B}\ }\textbf {\bibinfo {volume} {263-264}},\
  \bibinfo {pages} {421} (\bibinfo {year} {1999})}\BibitemShut {NoStop}%
\bibitem [{\citenamefont {Momma}\ and\ \citenamefont {Izumi}(2011)}]{momma}%
  \BibitemOpen
  \bibfield  {author} {\bibinfo {author} {\bibfnamefont {K.}~\bibnamefont
  {Momma}}\ and\ \bibinfo {author} {\bibfnamefont {F.}~\bibnamefont {Izumi}},\
  }\href@noop {} {\bibfield  {journal} {\bibinfo  {journal} {J. Appl.
  Crystallogr.}\ }\textbf {\bibinfo {volume} {44}},\ \bibinfo {pages} {1272}
  (\bibinfo {year} {2011})}\BibitemShut {NoStop}%
\end{thebibliography}%
\end{document}